\documentclass[11pt,twoside,A4]{article} 
\usepackage{times,fancyhdr}
\usepackage[dvips]{graphicx}
\usepackage{latexsym} 
\usepackage[affil-it]{authblk}
\usepackage{amsmath}
\usepackage{amssymb}
\usepackage{setspace}
\usepackage{hyperref}

\usepackage[top=4cm, bottom=4cm, left=4cm, right=4cm]{geometry}

\def\beq{\begin{equation}}
\def\eeq{\end{equation}}
\def\beqn{\begin{eqnarray}}
\def\eeqn{\end{eqnarray}}
\def\slash{/\hspace*{-2.2mm}}

\begin{document}
 
\title{Phenomenology of Space-time Imperfection II: Local Defects}
\author{S. Hossenfelder\thanks{hossi@nordita.org}} 
\affil{\small Nordita\\
KTH Royal Institute of Technology and Stockholm University\\
Roslagstullsbacken 23, SE-106 91 Stockholm, Sweden}
\date{}
\maketitle
%\centerline{DRAFT -- NOT FOR DISTRIBUTION}
\begin{abstract}
We propose a phenomenological model for the scattering of particles on space-time
defects in a treatment that maintains Lorentz-invariance on the average. The local
defects considered here cause a stochastic violation of momentum 
conservation. The scattering probability is parameterized in the
density of defects and the distribution of the momentum that a
particle can obtain when scattering on the defect. We identify the
most promising observable consequences and derive constraints
from existing data. 
\end{abstract}

\section{Introduction}

The phenomenology of quantum gravity proceeds by the development of models that parameterize
properties which the, still unknown, theory of quantum gravity 
might have. These phenomenological models are constructed for
the purpose of being testable by experiment and thereby guide
the development of the theory.
Among the best studied phenomenological consequences of quantum gravity are violations or deformations of Lorentz-invariance, additional spatial dimensions, and decoherence induced by quantum
fluctuations of space-time. Research in the area of 
quantum gravity phenomenology today encompasses a large variety of 
models that have been reviewed 
in \cite{AmelinoCamelia:2008qg,Hossenfelder:2009nu,Hossenfelder:2010zj}. 

A possible observable consequence of quantum gravity that has so far gotten 
little attention is the existence of space-time defects. If the seemingly smooth space-time that 
we experience is not fundamental but merely emergent from an underlying, non-geometric theory
 then we expect it to be imperfect -- it should have defects.
We will here study which observational consequences can be expected 
from such space-time imperfections of non-geometric origin.

Space-time defects are localized both in space and in time and therefore, in contrast to defects
in condensed-matter systems, do not have worldlines. There are two different
types of defects: Local defects and nonlocal defects. Local defects
respect the emergent locality of the space-time manifold. A particle that encounters a
local defect will scatter and change direction, but continue its world-line continuously. Nonlocal defects on the other hand do not respect the emergent locality of the 
space-time manifold. A particle that encounters a nonlocal defect continues its path in 
space-time elsewhere, but with the same momentum. The nonlocal
defect causes a translation in space-time, while the local defect causes a translation
in momentum-space.

The present paper is
the second part of a study of space-time defects and deals with
local defects. Nonlocal defects have been subject of the first part \cite{own}.
In principle a space-time defect 
could cause both, a change of position and momentum. But before making the situation
more complicated by combining these two effects, we will first study the two cases separately. In this paper, we 
develop a model for the local type of defects. Since Lorentz-invariance violation
is the probably most extensively studied area of Planck-scale physics \cite{Mattingly:2005re,Kostelecky:2008ts}, 
we will only consider the case where Lorentz-invariance is maintained on the average.
A different model for local space-time defects has 
recently been put forward in \cite{Schreck:2012pf}. We will briefly comment on the differences to this model
in the discussion. 

We use the unit convention $\hbar=c=1$. The signature of the metric is $(+,-,-,-)$.

\section{The distribution of defects}
\label{distribution}
 
To develop our model for local defects, we will here start with the simplest case in which the emergent background space-time
is flat Minkowski space, ie background curvature is not taken into account. This approximation will allow us to describe only
systems in which gravitational effects are negligible, but it is good enough for Earth-based laboratories and
interstellar propagation over intergalactic distances where curvature is weak and redshift can still be neglected. These
will be the cases we consider in section \ref{obs}.

The
only presently known probability distribution for points in Minkowski space that preserves
Lorentz-invariance on the average is the result of a Poisson process developed and studied in \cite{Dowker:2003hb, Bombelli:2006nm}.
With this distribution, the probability of finding  N points in a space-time volume $V$ is
\beqn
P_{\rm N}(V) = \frac{(\beta V)^{\rm N} \exp(-\beta V)}{{\rm N}!} \quad, \label{pois}
\eeqn
where $\beta$ is a constant space-time density. 

The 
average value of the number of points that one will find in some volume $V$ is then the expectation value
of the above distribution and given by
\beqn
\langle {\rm N}(V) \rangle = \sum_{{\rm N}=0}^\infty P_{\rm N}(V) {\rm N} = \beta V \quad.
\eeqn
The variance that quantifies the typical fluctuations around the mean is $\Delta {\rm N} \sim \sqrt{\beta V}$,
and the corresponding fluctuations in the density of points are $\Delta ({\rm N}/V) \sim \sqrt{\beta/V}$. 
In other words, the density fluctuations will be small for large volumes. 

We will
use the distribution (\ref{pois}) to seed the defects with an average density $\beta$. In
the following, we will not be concerned with fluctuations in the density as our aim here
is to first get a general understanding for the type and size of effects caused by local
defects and using the average will suffice for the present purpose. 
The probability is a density over space-time $\beta = L^{d+1}$, where $L$
is a length scale and $d$ is the number of spatial dimensions.  The ratio between
the fundamental length scale, that we take to be the Planck length $l_{\rm P}$, and
$L$ is $\epsilon = l_{\rm P}/L \ll 1$, ie the defects are sparse. Just exactly
how sparse is a question of experimental constraints that we will address in
section \ref{obs}.

\section{Kinematics of scattering on local defects}

The idea which we want to parameterize here is that local defects
are deviations from the smooth geometry of general relativity that
cause a violation of energy-momentum conservation.

The requirement of Lorentz-invariance restricts what the particle can
do when it encounters such a local defect. This restriction is more stringent than for
normal scattering processes because we have fewer quantities as input. We only
have the ingoing and outgoing momenta, whereas one normally has at least three
momenta involved in scattering, leading to the three invariant 
Mandelstam variables. 

\begin{figure}[ht]
\hspace*{1cm}\includegraphics[height=3.5cm]{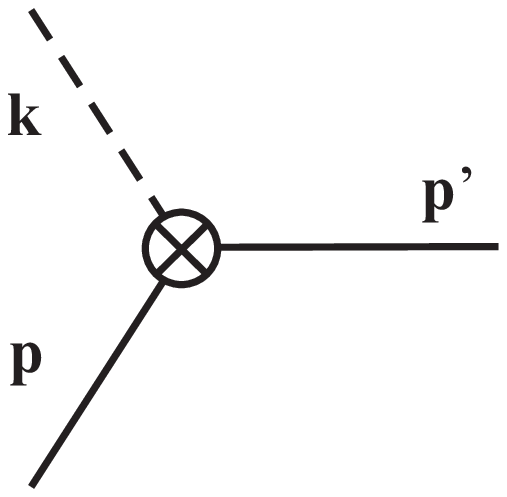} \hspace*{1.5cm} \includegraphics[height=3.5cm]{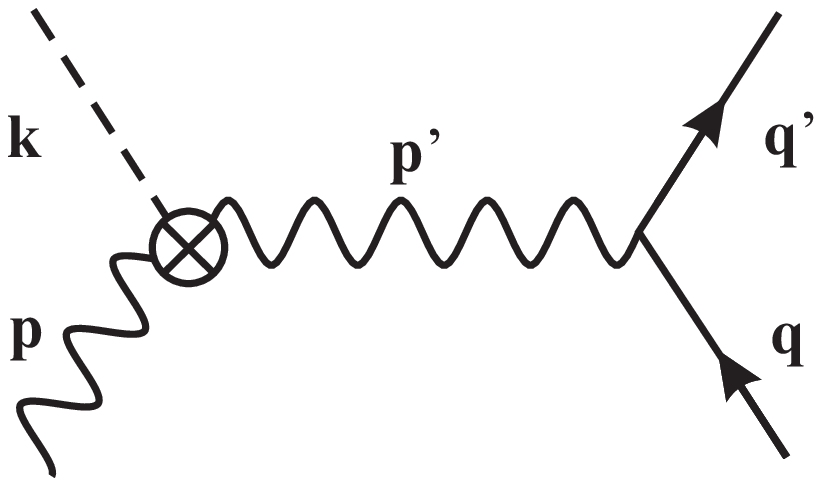}

\caption{Assignment of momentum notation. Left: Simple vertex for scattering on local
defect (dotted line). Right: Photon (wavy line) decays into a fermion pair (solid line) enabled by scattering on 
local defect (dotted line). \label{fig1}}
\end{figure}

Let us denote the momentum of the particle before
it encounters the defect with ${\bf p}$ and the momentum after it encountered the
defect with ${\bf p'}$, where boldfaced quantities denote four-vectors. Let us further formally assign the momentum ${\bf k} = {\bf p}' - {\bf p}$ to
the defect (see Figure \ref{fig1} left).
This 
assignment of momentum to the defect is a bookkeeping device that will
allow us to think in terms of normal scattering processes. The space-time 
defect itself does however not actually have a momentum; it instead causes a violation of
momentum conservation.

\subsection{Massless particles in 1+1 dimensions}

We will first
consider the case where the ingoing particle is right moving and massless, $m=0$.
Since ${\bf p}^2 = 0$, we
have two Lorentz-invariants left, ${\bf p}  \cdot  {\bf k} = M^2$ and ${\bf k}^2= a^2 M^2$ (thus ${\bf p}'^2 = (a^2 + 2) M^2$),
where $m$ and $M$ have dimension mass, and $a$ is a dimensionless
parameter that we expect to be of order one.  It is henceforth
assumed that $a$ and $M$ are real-valued to avoid that the mass of the
outgoing particle can become tachyonic. 

%..........................................................................
%..........................................................................

We want to quantify now the probability ${\cal P}$ for what
will happen when the particle encounters the defect. Lorentz-invariance requires
${\cal P}$ to be a function solely of $a$ and $M$, and our expectation is that it
is actually a function of $a^2$ and $M^2$.  The
condition ${\bf p} \cdot {\bf k} = M^2$ selects a cut in the hyperboloid defined by
${\bf k}^2=a^2 M^2$. In 1+1 dimensions, this specifies ${\bf k}$ completely. The
higher dimensional case brings additional complications that we will come to
in section \ref{4d}.

When we leave behind the classical particle and think about quantum particles,
it does not seem to make much sense to have a particle scatter off a point
to obtain a distinct momentum from that point. We should instead take into
account that the point has a position uncertainty and the momentum it
transfers will inevitably have an uncertainty too. That is 
the distribution ${\cal P}(M^2, a^2) $ will not encode distinct
values of $a$ and $M$ but these variables will have some spread to them. 
We will
quantify their distribution only roughly by means of the average values $\langle a^2 \rangle,
\langle M^2 \rangle$ and variances $\Delta (a^2), \Delta (M^2)$. For
the sake of readability, we will in the following write the variances simply as  $\Delta a^2, \Delta M^2$.

Let us assume that the massless particle moves into the positive $x_1$-direction, and
denote $p^\nu = (E,E)$ and $k_\nu = (k_0, k_1)$, where $k_1 = \pm \sqrt{k_0^2 - M^2/a^2})$. Then the
requirement ${\bf p} \cdot {\bf k} = M^2$ leads to
\beqn
E =  \frac{k_0}{a^2} \mp \sqrt{\frac{k_0^2}{a^4} - \frac{M^2}{a^2}}  \quad.
\eeqn
Or, solving for ${\bf k}$ instead, one finds
\beqn
k_0 = \frac{1}{2} \left(\frac{M^2}{E} + a^2 E \right) \quad,\quad k_1 = \frac{1}{2}  \left(\frac{M^2}{E} - a^2 E \right) \quad. \label{kx}
\eeqn
This means in particular there is no threshold for the particle scattering; massless particles
can scatter even if they have a very small energy.

Though the assignment of momentum to the defect is just a bookkeeping
device it is instructive to make a Fourier
transformation of the momentum distribution ${\cal P}(M^2, a^2)$ . To that end, we
consider the distribution to be Gaussian
\beqn
{\cal P}(M^2, a^2) = \exp\left(- \frac{(M^2 -\langle M^2 \rangle)^2}{(2 \Delta M^2)^2} - \frac{(a^2 - \langle a^2 \rangle)^2}{(2 \Delta a^2)^2} \right) \left(2 \pi \sqrt{\Delta M^2 \Delta a^2} \right)^{-1} \quad, 
\eeqn
where we have placed the defect at the origin. We 
now want to compute the space-time distribution
\beqn
P(t,x) = \int d (M^2) d (a^2 ) {\cal P} (M^2, a^2) e^{ - {\rm i} (k_0 x^0 + k_1 x^1)} \quad. \label{Ftx}
\eeqn

Since we are dealing with massless particles, it will be handy to introduce
the light-cone momenta and $k^\pm = k_\mp = ( k^0 \pm k^1 )/\sqrt{2}$ which are
\beqn
k_+ = \frac{1}{\sqrt{2}} \frac{M^2}{E} \quad, \quad k_- = \frac{1}{\sqrt{2}}  a^2 E \quad, \label{kpm}
\eeqn
and $p_+ = E$, $p_- =0$. We then rewrite the integral into lightcone coordinates and obtain
\beqn
P(x^+, x^-) &=&  \int d k_+ d k_-   {\cal P} (k_+, k_-) e^{ - {\rm i} (k_+ x^+ + k_- x^-)} %\quad,
\eeqn
(the Jacobian is equal to one) where $x^\pm = (x^0 \pm x^1)/\sqrt{2}$  and
\beqn
{\cal P} (k_+, k_-) =  \exp\left( - \frac{E^2(k_+ - \langle k_+ \rangle)^2}{2 (\Delta M^2)^2} - \frac{2 (k_- - \langle k_- \rangle)^2}{2 (E \Delta a^2)^2} \right) \left(2 \pi \sqrt{\Delta M^2 \Delta a^2} \right)^{-1} \quad. \label{comthis}
\eeqn
Here we have introduced the average values $\langle k_+ \rangle = \langle M^2 \rangle / (\sqrt{2} E)$ and $\langle k_- \rangle = \langle a^2 \rangle E/\sqrt{2}$. The Fourier-transformation thus yields a Gaussian distribution in light-cone coordinates
\beqn
P(x^+, x^-) &=& \exp\left(\frac{(x^+)^2}{(2 \sigma^+)^2} + \frac{(x^-)^2}{(2 \sigma^-)^2}\right) 
\frac{e^{{\rm i} (\langle k_+ \rangle x^+ + \langle k_- \rangle x^-)}}{2 \pi \sqrt{\sigma^+ \sigma^-}}
\quad,
\eeqn
with widths
\beqn
\sigma^+ = \frac{\sqrt{2} E}{\Delta M^2} \quad, \quad \sigma^- = \frac{\sqrt{2}}{E \Delta a^2} \quad. \label{sigmas}
\eeqn
We see that the typical space-time patch covered in the direction of propagation
\beqn
\sigma^+ \sigma^- = 2 \left(\Delta M^2 \Delta a^2 \right)^{-1} 
\eeqn
has a Lorentz-invariant volume independent of $E$, though it will deform under boosts that red- or blueshift $E$ as one sees from Eqs. (\ref{sigmas}).

To make sense of the scarcity of local defects, the typical size associated with the
defect should be much smaller than the typical volume in which to find one
defect, ie $L^2 \Delta M^2 \Delta a^2 \gg 1$. Strictly speaking, the above treatment is valid only for an incident plane wave with exactly
defined momentum $E$. If the incoming particle has an energy spread, $\Delta p_+ \neq 0$, this will
contribute to $\Delta k_\pm$ via the relations (\ref{kpm}). The plane wave approximation
is good only as long as $\langle a^2 \rangle \Delta p_+ \ll \langle p_+ \rangle \Delta a^2$ and $\langle M^2 \rangle \Delta p_+ \ll p_+ \Delta M^2$. 
We will in the following assume that this approximation is valid.

\subsection{Massive particles in 1+1 dimensions}

If the incoming particle has a mass, $m > 0$, we have an additional dimensionful parameter
to generalize our requirements on ${\bf k}^2$ and ${\bf p} \cdot {\bf k}$. To find a
suitable way to treat massive particles, we first note that in the ultrarelativistc 
limit the relation for ${\bf k}^2$ should reproduce the $k_+$ from the massless
case. Thus, we will leave the requirement ${\bf k}^2 = a^2 M^2$ the same
for the massive case as it was for the massless case. We further note that in
the limit of small energies $k_0$ and $k_1$ from the massless case tend towards
the same value. We therefore expect that for a massive particle in the restframe $k_0$
and $k_1$ have the same value. 

We thus make the ansatz ${\bf k} \cdot {\bf p} = M^2 + a^2 m^2/4$, which in the
restframe of the massive particle leads to
\beqn
k_0 = \frac{M^2}{m} + \frac{a^2 m}{4} \quad, \quad k_1 = \pm \Big| \frac{M^2}{m} - \frac{a^2 m}{4} \Big| \quad. \label{kmass}
\eeqn
If the incoming particle has a mass, $m$, then the defect will increase its mass
to ${\bf p}'^2 = M^2(a +2) + m^2 (1 + a^2/2) $
and can thus result in a virtual particle as in the massless case. 

This leads us to assign the following
Gaussian distribution to the defect 
\beqn
{\cal P}(M^2, a^2) = \exp \left( - \frac{(M^2 - \langle M \rangle^2)^2 }{ (2 \Delta_{ m} M^2)^2}  - \frac{(a^2 - \langle a^2 \rangle)^2}{ (2 \Delta_{m} a^2)^2} \right) \left(2 \pi \sqrt{\Delta_m M^2 \Delta_m a^2} \right)^{-1},
\eeqn
where the index $m$ on the variance stands for the mass of the particle and indicates that this distribution is a priori not the same
as in the massless case. We proceed as previously and rewrite this into momentum space
\beqn
{\cal P}(k_0, k_1) &=& \exp \left( - \frac{(k_0 - \langle k_0 \rangle + |k_1| - \langle |k_1| \rangle)^2 m^2}{2 (\Delta_{m} M^2)^2} \right) \times \nonumber \\
&& \hspace*{-1cm}\exp \left( - \frac{(k_0 - \langle k_0 \rangle - | k_1| + \langle |k_1| \rangle)^2}{2 m^2 (\Delta_{m} a^2)^2} \right) \left(2 \pi \sqrt{\Delta_m M^2 \Delta_m a^2} \right)^{-1} ,
\eeqn
where $\langle k_0 \rangle$ and $\langle k_1 \rangle$ are the mean values of $k_1$ and $k2$ in (\ref{kmass}).
Since we expect the variances of $k_0$ and $k_1$ to be the same in the restframe we set $\Delta_{m} a^2 m = \Delta_m M^2  /m  = \Delta_m k$, which yields
\beqn
{\cal P}({\bf k}) = \exp \left( - \frac{(k_0 - \langle k_0 \rangle)^2 + k_1^2}{(2 \Delta_{m} k)^2} \right) (2 \pi \Delta_m k)^{-1}  \quad. \label{genthis}
\eeqn
In space-time the defect in the restframe of the incident particle is thus described by the Fourier-transformation
\beqn
P({\bf x}) = \exp \left( - \frac{(x_0^2 + x_1^2)}{( 2 \Delta_{m} x)^2} \right)  e^{i \langle k_0 \rangle x^0 }  (2 \pi \Delta_m x)^{-1} \quad,
\eeqn
where $\Delta_m x = m/\Delta_m M^2$. 

Since the space-time volume is Lorentz-invariant, we know then that the typical volume 
occupied by the defect is $(m/\Delta_m M^2)^2$. If we want this volume to
be the same as in the massless case, we can identify $(\Delta_m M^2)^2 = m^2 \Delta M^2 \Delta a^2$ and 
$(\Delta_m a^2)^2 =  (\Delta M^2 \Delta a^2)/m^2$.

\subsection{Massive and massless particles in 3+1 dimensions}
\label{4d}

The reason we were able to construct the above Lorentz-invariant and normalizable
distributions even though the Lorentz-group is not
compact is that we have not treated the momentum distribution assigned to the
defect as
internal, but as dependent on the incident particle. Then, one can use the
momentum vector of the particle that scatters as reference without affecting 
observer-independence of the outcome. 

If we add two additional spatial dimensions, momentum space becomes larger and the two requirements on
${\bf k}$ will no longer suffice to pick out a specific momentum that the defect transfers. The
degeneracy in ${\bf k}$ is due to the subgroup of the Lorentz-group
that leaves the momentum vector ${\bf p}$ invariant, known
as the little group of the particle. If we have only ${\bf p}$ as reference,
then in each reference frame in which
${\bf p}$ has the same components, the distribution for ${\bf k}$
must be the same. 

In 3+1 dimensions, the little group
of a massive particle is  $SO(3)$ and compact, so the treatment is
straight-forward because we can use a uniform distribution
over the additional degrees of freedom. We add the 
requirement that the restframe of the particle
remains the same on the average, and the distribution is spherically symmetric
for the spatial components of ${\bf k}$ in the particle's restframe.

This means 
\beqn
k_0 = \frac{M^2}{m} + \frac{a^2 m}{4} \quad, \quad k = \Big| \frac{M^2}{m} - \frac{a^2 m}{4} \Big| \quad,
\eeqn
where $k = |\vec k |$ and $\vec k = (k_1, k_2, k_3)$ and the 1+1 dimensional distribution generalizes to (compare to 
Eq. (\ref{genthis})) 
\beqn
{\cal P}({\bf k}) = \exp \left( - \frac{(k_0^2 + k^2)}{( 2 \Delta_{m} k)^2} \right) (2 \pi \Delta_m k)^{-2} \quad.
\eeqn
In space-time, the defect is thus described by the (four dimensional) Fourier-transformation
\beqn
P({\bf x}) = \exp \left( - \frac{(x_0^2 + x^2)}{( 2 \Delta_{m} x) ^2} \right) (2 \pi \Delta_m x)^{-2} \quad,
\eeqn
where $\Delta_m x = m/\Delta_m M^2$ as before and the typical volume 
occupied by the defect is $(m/\Delta_m M^2)^4$. This distribution is
invariant under the little group of the incident massive particle that scatters on
the defect and thus all observers
who measure the particle must agree on the scattering result, as desired.

The little group for a massless particle it is $ISO(2)$, 
the symmetry group of the 2-dimensional Euclidean plane. It consists
of rotations in the $x_2 - x_3$ plane and combinations of boosts and
rotations\footnote{The existence of these additional transformations goes back to the
same property of the 3+1 dimensional Lorentz-group that gives rise to the Thomas-Wigner rotation.}. For an explicit construction, see eg \cite{Weinberg}. $ISO(2)$ is not
a semi-simple group and not compact. The rotational part into direction
$x_2-x_3$ is unproblematic, but the remaining transformations
have a continuous set of two parameters corresponding to an infinite
number of reference frames in which observers will all measure
the same components of the incident massless particle's four-momentum. 

Since  ${\bf k}$ is then not
uniquely defined but only up to the transformations in the little group of the
incident particle, without further input, Lorentz-invariance would require that we have to treat all possible
solutions equally. Because the parameter space is not compact, for a massless particle 
this would then entail a uniform distribution over two unbounded parameters which
is not normalizable. 

However, we deal with uniform distributions over non-compact spaces 
every time we use plane waves. It thus seems likely that the problem has
the same origin and can be dealt with the same way, namely by taking
into account that in reality the incoming particle is spread out only over
a finite volume and has a non-zero momentum spread.

Concretely, the elements of the photon's little group are besides the
rotations in the $x_2 - x_3$ plane of the form
\beqn
L^\mu_{\; \nu} = \left( 
\begin{array}{cccc}
1 + \xi & -\xi & \chi & \kappa \\
\xi & 1-\xi & \chi & \kappa \\
\chi & - \chi & 1 &0 \\
\kappa & -\kappa & 0 & 1 
\end{array}
\right) \quad,
\eeqn
where $2 \xi = \chi^2 + \kappa^2$. One convinces oneself readily that indeed $L^\mu_{\; \nu} p^\nu = p^\mu$. 

The solution for ${\bf k}$ from the 1+1 dimensional case is still a solution to the
requirements on ${\bf k} \cdot {\bf p}$ and ${\bf k}^2$  (with two zero entries added), but under a transformation of the above photon's little
group element, one obtains 
\beqn
k'^\nu = L^\mu_{\; \nu} k^\mu = k^\nu + \frac{M^2}{E} v^\nu \quad, \label{ktrafo}
\eeqn
where $v^\nu = (\xi, \xi, \chi, \kappa)$. The probability distribution for ${\cal P}({\bf k})$ in
four dimensions then should be independent of $\kappa$ and $\chi$, where $\kappa$
and $\chi$ could be expressed in terms of $k'_2$ and $k'_3$ as $\chi = k'_2 E/M^2$ and $\kappa = k'_3 E/M^2$.

To address the issue of normalizing the distribution, we now take into account that the incident particle's wave-function has a finite spread
in $p_2$ and $p_3$ direction\footnote{Note that this spread is not constant over time.}, which we will label $\Delta p_2$ and $\Delta p_3$. The
mean momentum vector we will denote $\langle {\bf p} \rangle = (\langle E \rangle, \langle E \rangle, 0, 0)$. We
would like to preserve the rotational symmetry, and so will assume $\Delta p_2 = \Delta p_3 =: \Delta p_\perp$.
These relevant point is that these extensions in direction $x_2$ and $x_3$ are perpendicular to the photon's wave-vector and not
invariant under the little group. Therefore, they provide us with additional information about the incident particle
that we can use to put bounds on the integration over $\kappa$ and $\chi$, or (dropping the primes) on $k_2$ and $k_3$ respectively.
We do this by assuming that the width of the defect in momentum space has 
$\Delta k_2 = \Delta k_3 =: \Delta k_\perp = \Delta M^2/\Delta p_\perp$ and so
(compare to Eq. (\ref{comthis}))
\beqn
{\cal P} ({\bf k}) =  \exp\left( - \frac{\langle E \rangle ^2 k_+^2}{(\Delta M^2)^2} - \frac{(k_- - \langle k_- \rangle)^2}{(\langle E \rangle \Delta a^2)^2}
-\frac{k_\perp^2 }{2 \Delta k_\perp^2} 
 \right) \left(4 \pi^2 \Delta k_\perp \sqrt{\Delta M^2 \Delta a^2} \right)^{-1} \label{massless4d}
\eeqn
where $k_\perp^2 =k_2^2 + k_3^2$. 

In (\ref{massless4d}) we have set $\langle M^2 \rangle = 0$. In this case, the average of ${\bf k}$ is parallel to the average of ${\bf p}$. Any
other choice will, in the plane wave limit (when $\langle E \rangle = E$), not be invariant under the little group as one sees from
Eq (\ref{ktrafo}). It is thus henceforth assumed 
that $\langle M^2 \rangle =0$ and thus $\langle k_+ \rangle = 0$. In the
plane-wave limit where $\Delta p_\perp \to 0$, one then has $\Delta k_\perp \to \infty$ and the distribution of ${\bf k}$ will 
go to the uniform distribution over the little group.

This construction for massless particles in 3+1 dimensions is not 
invariant under the little group of the mean momentum of the incoming photon wave packet. But this
is unnecessary because the incident
particle, when it has a finite width, has the rotational symmetry in $x_2 -x_3$
as the only remaining symmetry. Thus, observer independence is preserved and the distribution is normalizable. In the
plane-wave limit, the components of ${\bf k}$ can become arbitrarily large
and then the defect can in principle transfer an arbitrarily large momentum to the 
particle. In reality, this momentum transfer will however be bounded because the incident
particle has a finite width. We have essentially identified the non-compact part
of $ISO(2)$ with the non-compactness of the plane wave.

In summary, we have seen that Lorentz-invariance proves to be surprisingly
restrictive on the possible scattering outcomes. We have dealt with the 
non-compactness of the Lorentz-group by using measurable properties
of the incident particle to reduce the symmetry of the momentum
non-conservation mediated by the defect while preserving observer-independence. 

\section{Dynamics of scattering on local defects}

This now leads us to ask what we can say about scattering amplitudes. 
For concreteness, we assume that the defect makes
itself noticeable in the covariant derivative 
since the local defect represents a non-geometric 
inhomogeneity that cannot be accounted for by the normal covariant
derivative. 

The local defect thus appears in the Lagrangian together with the derivative terms and
gauge fields and can be implemented by replacing the normal gauge-invariant
derivative $D = \partial + e A$ with $D = \partial + e A + \tilde g \partial P$, where $P$ is 
the previously introduced Fourier-transform of the momentum distribution ${\cal P}$,
 and the derivative $\partial P$ is proportional to ${\bf p}'- {\bf p} = {\bf k}$. $e$ is some Standard Model coupling constant
(not necessarily that of {\sc QED} though this will be the case we consider later)
and $\tilde g$ is the coupling constant for the defects. Since the mass dimension of
$[P] = [E^2]$, the dimension of $[\tilde g] = [E^{-2}]$, and since we have only one
scale of dimension mass at our hands, we use $\tilde g = 1/\Delta M^2$. 

The defects are sprinkled over space-time according to the Poisson
process described in section \ref{distribution}, but since we are
considering the homogenous case we will not treat the defects as
individual events. We will instead replace the many single defects
by a  field $P$ with which the
standard model fields have a small interaction probability. That
is, instead of interacting with defects in rare places, the particle
interacts with the defect-field anywhere but with low probability. 

In this
limit, instead of having a sum over many defects, the probability to
interact with 
the defect-field $P$ is suppressed by the volume of the defect over
the typical volume in which to find a defect, ie 
\beqn
\tilde g^2 = \frac{1}{(\Delta M^2)^2} \frac{\sigma^+ \sigma^-}{\Delta k_\perp^2 L^4} \quad.
\eeqn

This is a similar approximation as we have made in \cite{own} in that it is
implicitly assumed here that scattering on defects happens often enough so
that we can effectively describe it as a homogeneously occuring process.
Concretely this means that this approximation is only good if the 
space-time volume swept out by a particle's wave-function is large enough
to contain a large number ($\gg 1$) of defects. In this sense it is an
effective description, not unlike in-medium propagation of photons, except
that here we maintain Lorentz-invariance and thus the effects (have to) scale
differently.

We note however that not only is the coupling constant dimensionful, but also
goes to zero in the plane wave-limit, when $\Delta p_\perp \to 0$ and $\Delta k_\perp \to \infty$. 
This behavior is an artifact of the normalization and
dimension of the field $P$ which contains a prefactor of $\Delta k_\perp$. To avoid
having to deal with quantities that diverge in the plane wave limit, we will thus
shift this normalization factor from $P$ into the coupling constant, so that both
become dimensionless. We therefore define
\beqn
\hat P ({\bf x}) &:=& \frac{P({\bf x})}{\Delta k_\perp \sqrt{\Delta M^2}} \quad, \quad \hat {\cal P} ({\bf k}) 
:= \frac{{\cal P}({\bf k})}{\Delta k_\perp \sqrt{\Delta M^2}} \quad, \\
\quad g &:=& \tilde g \Delta k_\perp \sqrt{\Delta M^2} = \frac{1}{\Delta M^2} \frac{1}{\sqrt{\Delta a^2}} \frac{1}{L^2} \quad. 
\eeqn
With this definition, the coupling to standard model fields is then $D = \partial + e A +  g \partial \hat P$ and
$g$ is dimensionless and remains finite in the plane wave limit.

We will further in the following assume that the defects do not carry any standard model charges 
and do not change the type of the ingoing particle, ie the scattering is
entirely elastic. In principle we could have a defect that changes 
not only the momentum but also the spin of the particle, but we will not
consider this possibility here. 

That the assignment of $\hat P({\bf x})$ to the defect is a bookkeeping device rather
than the introduction of a new field is somewhat hidden in this notation,
which looks like we have introduced a new field. 
That $\hat P({\bf x})$ is not actually an independent quantity can be seen 
most clearly from Eqs (\ref{Ftx}) and (\ref{kx}).
The
momentum ${\bf k}$ that is assigned to the defect is a derived
quantity from the particle's incoming momentum ${\bf p}$, and $\hat P$ is
thus an operator acting on the same field that the gauge-covariant derivative acts
on.  

From this function of $\hat P({\bf x})$ it is also clear that the scattering on
the local defect, in the form that we have introduced it here, will
break gauge invariance. This might not come as a surprise since the
simplest Lorentz-invariant change to the dynamics of a massless gauge
field is adding a mass term. Here, the underlying reason for the breaking of
gauge invariance is that the defect itself is not gauged, ie the ${\bf k}$ that is
derived from the incoming momentum ${\bf p}$ does not respect the
gauge of the incoming particle. 

One could fix gauge invariance by appropriately adding the gauge field to
${\bf k}$ from the beginning on, but then one would obtain a nasty integral with 
gauge fields in exponents. Because of this complication together 
with the breaking of gauge invariance being not unexpected, we will
accept that gauge invariance is broken. 
In the abelian case the additional terms for coupling a fermion field and
its gauge-field to the defect are then
of the form $g \Psi^\dag (\slash \partial \hat P) \Psi$, $e^2 g \partial A (\partial \hat P) A$, $e^2 g^2 (\partial \hat P) A (\partial \hat P) A$.

With this prescription one can now calculate the scattering
amplitudes for processes of interest. For each 
amplitude where one of the incoming particles previously
scattered on the defect, one gets an additional integral
over $d (a^2) d (M^2) d k_2 d k_3 $, or $d k_+ d k_- d^2 k_\perp = d^4 k$ respectively. From
this one obtains a cross-section or decay rate as usual.

Before we turn towards some examples, let us make a general observation.
Since $\langle M^2 \rangle = 0$ and ${\cal P}$ does not vanish for $M^2 = 0$
if we assume a Gaussian distribution, it seems that a massless on-shell particle could
remain on-shell and the momentum of the outgoing particle ${\bf p}'$ would then be a multiple
of the momentum of the ingoing particle ${\bf p}$. However, unless the incident particle is virtual, 
the vertex factors all vanish because there is then no non-zero contraction of the
momentum ${\bf k}$ and the massless particle's momentum or its 
transverse polarization tensor. Thus, if the incoming massless particle is real, the
outgoing massless particle will necessarily be virtual. 

For the massive particle, when $M^2 = 0$ the
outgoing momentum must fulfill ${\bf p}'^2 = (1+a^2)m^2$ and thus the
particle is virtual unless also $a^2 = 0$ in which case the particle only 
receives a transverse momentum from the defect that is equally distributed
in its restframe. Alas, if $M^2 = a^2 = 0$, the particle did not acquire
kinetic energy and thus a non-zero transverse momentum would violate the
mass-shell condition.  In other words, the massive particle too must
be off-shell after changing its momentum by scattering on the defect.

Since the defects are sparse or the coupling constant is small respectively, we
expect effects to be small and noticeable primarily for long-lived particles.

\section{Constraints from observables}
\label{obs}

We will in the following make the assumption that $\langle a^2  \rangle \sim \Delta a^2 \sim 1$. 
Then we are left with two parameters, $\Delta M^2$  and $L$. (For some discussion on this and
the other assumptions,
see next section.)

Let us start with some general remarks. Since we are looking for constraints on (normally) stable particles that propagate over long times
or distances respectively, we will focus on the {\sc QED} sector of the standard model and
on particles that we observe coming from distant astrophysical sources. This
makes in particular the photon a sensitive probe for the presence of local
defects.

If a photon scatters on the defect and
becomes a virtual photon with mass $\sim \sqrt{3} M$, it can only decay into a fermion
pair if $\sqrt{3} M$ is larger than twice the mass of the fermion. If $\sqrt{3} M$ is smaller
than twice the electron mass, this leaves decay into a neutrino and antineutrino
as the only option, which would necessitate a higher-order electroweak
process and dramatically lower the cross-section. If $M$ is smaller than 
even the lightest neutrinos, the only option left for the virtual photon is 
to scatter on another defect.

The phenomenology thus depends significantly on the value of $\Delta M^2$
because it determines the relevance of possible decay channels through
the typical range in the probability distribution ${\cal P}$. 
The is always some possibility for $M^2$ or $a^2$ to take on very small or very large values,
but the probability for this to happen is highly suppressed if $M^2$ is many orders of
magnitude beyond $\Delta M^2$.  For the
rest of this section, we will only estimate the orders of magnitude for typical values of $M$,
and will therefore from now on just write $M^2$ for $\Delta M^2$ 
and
omit factors of order one.

In order to get a grip on the phenomenology, let us thus 
identify and focus on the parameter range that seems most interesting.

\begin{figure}[ht]
%\centering 
\includegraphics[height=2.8cm]{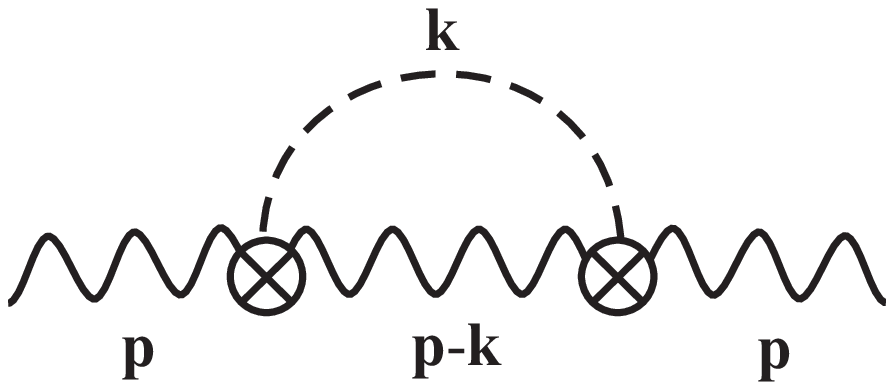} \hspace*{1cm} \includegraphics[height=3.2cm]{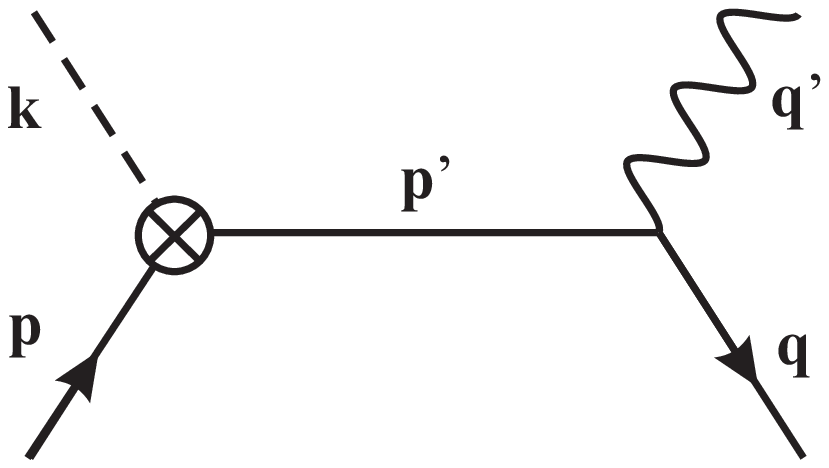}

\caption{Left: Contribution to photon mass from twice scattering on a defect. Right: Vacuum Cherenkov radiation, enabled by
scattering on a defect. \label{fig2}}
\end{figure}

As mentioned previously, to make sense of the defects, we expect their typical
volume to be small in comparison to the typical distance between the defects
which is given by the density $\beta$. Or, in other words, we expect the effective
coupling constant in the homogeneous limit, $g$, to be much smaller than one. 
Since we have already introduced one
small parameter ${\epsilon} = l_{\rm P}/L$ that is the (fourth root of) the
density of the defects over a Planck density, the range
$ M^2 L^2 \sim 1/{\epsilon} \gg 1$ is of particular
interest. If furthermore $L$ is in the same range as the length scale associated with
the cosmological constant (a tenth of a millimeter or so), then 
$M$ is approximately in the TeV range. This is the parameter
range that we will focus on in the following\footnote{The energy and the length scales are not necessarily the same as those
parameterizing the effects of nonlocal defects in \cite{own}.}. In this parameter range, there is then
no problem for the virtual photon to decay into fermions, and there are three processes of main interest:
\begin{enumerate}
\item {\bf Photon decay:} The photon can make a
vacuum-decay into a pair of electrons via
a diagram as shown in Fig \ref{fig1}, right. This results in
a finite photon lifetime, and leads to electron-positron pair production. 
This process is similar to pair production in the presence of an
atomic nucleus in standard {\sc QED}.
\item {\bf  Photon mass:} The photon acquires an effective mass by scattering
into a virtual photon on a first defect, and then converting back into
a real photon by interacting with the defect a second time. This process is
depicted in Fig \ref{fig2}, left.
\item {\bf Vacuum Cherenkov radiation:} An electron can emit
a (real) photon after scattering on a defect, shown in Fig \ref{fig2}, right.
This process is similar to Bremsstrahlung in standard {\sc QED}.
\end{enumerate}

%..........................................................................
%..........................................................................

\subsection{Constraints on photon mass}

The contribution to the photon mass come from terms of the form $e^2 g^2 (\partial \hat P) A (\partial \hat P) A$.
Since the derivative in $\partial_\nu \hat P$ by definition just creates the defect's momentum ${\bf k}$ and
${\bf k}^2 = M^2$, this term produces a photon mass of approximately $m_\gamma \approx g M$. The best current
constraint on the photon mass is $m_\gamma \leq 10^{-18}$eV \cite{Goldhaber:2008xy}. This leads to
the bound
\beqn
M L^2 \geq 10^{-12} {\rm m} \quad.
\eeqn

\subsection{Photon lifetime}

If photons of the Cosmic Microwave Background ({\sc CMB}) decay before reaching us, this would
lead to deviations from the thermality of the CMB spectrum \cite{Heeck:2013cfa}. To obtain a
constraint, let us therefore
estimate the decay rate of photons from scattering on defects, induced by the
process depicted in Fig \ref{fig1}, right. The amplitude for this process has the form
\beqn
{\cal M} &\sim& - {\rm i} \left( \frac{e g}{\Delta k_+ \Delta k_- \Delta k_\perp^2} \right) \left( \frac{e_\mu}{\sqrt{E}} \right) u({\bf q}) u({\bf q}') \times \nonumber \\
&&\int d^4 p' \int d^4 k \frac{\eta_{\mu \nu}}{{\bf p}'^2} \gamma^\nu \delta({\bf p}' - {\bf q} - {\bf q}') {\bf p}'^2 \delta({\bf p}' - {\bf p} - {\bf k}) {\cal \hat P}({\bf k}) \quad, \label{amp}
\eeqn
where we have omitted some factors of $2 \pi$ and $\sqrt{2}$. 

In the order displayed, the amplitude (\ref{amp}) is composed of
the coupling constant for the first and second vertex, the normalization of ${\cal \hat P}({\bf k})$, the polarization tensor of the incoming
photon $e_\mu$ with (dimensionful) normalization, the spinor wave-functions of the outgoing electron and positron $u({\bf q})$ and $u({\bf q}')$ with
normalization omitted, the photon propagator, the first vertex and the second vertex, multiplied with the probability
distribution and integrated over the momentum of the virtual photon and that of the defect.

Omitting the polarization and spinor structure, we can perform the integral over ${\bf q}$ and estimate the integral over ${\cal \hat P}({\bf k})$
by evaluating the integrant at one standard deviation (times the variances, which cancel with the prefactor). This gives 
\beqn
{\cal M} &\sim& - {\rm i}  \frac{e g}{\sqrt{E}}  \delta({\bf p}' + {\bf k} - {\bf q} - {\bf q}')  \quad, 
\eeqn
where ${\bf k}$ is given by Eqs. \ref{kx} (recall that we replaced $\Delta M^2$ with $M^2$). From
this we obtain an estimate for the decay rate
\beqn
\Gamma (\gamma \to e^+ e^-) \approx E \alpha g^2 \quad.
\eeqn
The photon half-life time $\tau_\gamma$ is
thus
\beqn
\tau_\gamma \approx \frac{L^4 M^4}{\alpha E_0 z_0} \int_0^{z_0} d z a(z) \quad,
\eeqn
where $E_0$ is the photon energy at the present time, $z_0 \approx 1100$ the redshift
at the time of production of the photon, and $a(z)$ is the redshift-dependent
scale factor. With
 $E_0 \sim 10^{-2}$eV for a typical {\sc CMB} photon, the requirement that no more 
than about $10^{-4}$ {\sc CMB} photons should have decayed at the present time leads
to $\tau_\gamma \geq 10^{21}$s and
\beqn
L M \geq 10^{8} \quad.
\eeqn

This constraint however assumes that the density of defects remains constant in time,
the case that was also considered in \cite{PrescodWeinstein:2009wq}. If the density of 
defects dilutes, then it would have been higher in
the past, thereby decreasing the average decay time and tightening the constraint. It would need a more sophisticated model
for the generation of defects to know how the density evolves in time. However, if one makes the
ad-hoc assumption that $L(z) \sim L_0 a(z)$, $M = M_0$, then the constraint (on $L_0 M_0$) is by a
factor of about $10^3$ stronger.

\subsection{Cosmological vacuum opacity}

Besides affecting the {\sc CMB} spectrum, decaying photons will furthermore generally diminish the luminosity of faraway sources
while at the same time not changing the redshift. Constraints on such a
cosmological vacuum opacity have recently been summarized in \cite{Jimenez:2013mga}.
However, the constraints from the {\sc CMB} are stronger than the
constraints from emission of distant astrophysical light sources, owing
to the long travel-time of {\sc CMB} photons and the excellent precision by which their
spectrum has been measured. 

Because the photon sources in this case are localized however, the astrophysical constraints on
vacuum opacity would be
interesting to look for effects of inhomogeneities that might be difficult
to extract from the {\sc CMB} data. Since here we do not consider
inhomogeneities we will not quantify this constraint, but just mention that
it could prove interesting in the more general case.

\subsection{Heating of the CMB}

As one expects from our previous discussion, the total decay rate of photons is finite due to the
normalization procedure with a finite width of the ${\bf k}$ distribution. If
we take the plane wave limit, the total cross-section remains unmodified by construction but
the differential cross section now includes arbitrarily high momenta. The typical
momentum of the outgoing electron is then of the order $\Delta M^2 \Delta x$, where $\Delta x$ is the
width of the incident particle's wave-packet. If we assume $\Delta x \sim 1/E_0$ 
(note that this is not an observer-independent statement), the
momentum will be of the order of the Planck mass in the restframe
where $\Delta x$ takes on this value (that we identify with the {\sc CMB} or Earth
restframe, the distinction does not matter for our estimate). 

An electron of that high an energy
however has a very short lifetime because it will undergo inverse Compton scattering on {\sc CMB}
photons.  It has a huge $\gamma$-factor of about $10^{22}$ and thus an average mean free path $l$ of about \cite{Beckmann}
\beqn
l \sim  \frac{10^{-12}}{\gamma} {\rm light years} \sim 10^{-4} {\rm fm} \quad,
\eeqn
which means we'll never see it; it will just deposit its energy into the {\sc CMB}. At
such high energies, even the outgoing photon will have a short mean free path
because it scatters on other {\sc CMB} photons via box diagrams. 

Effectively, the two processes
of photon decay and vacuum Cherenkov radiation therefore just heat up the {\sc CMB}. Or
rather, they prevent it from cooling. Since the universe contains more free
photons than electrons, photon decay is the more relevant of these
processes. This then allows us to make the following rough estimate. The energy that is deposited into
the {\sc CMB} by the photons' scattering on defects 
should not significantly raise the {\sc CMB} temperature. This means that the typical
probability for the photon decay to happen, $g^2 \alpha$,
should be less than the ratio of the initial photon's energy over the outgoing photon's
energy $\Delta M^2/E_0$. This leads to the bound
\beqn
L^2 M \geq 10^{-2} {\rm m} \quad,
\eeqn
which is considerably stronger than the bound from photon masses.

\subsection{Summary of constraints}

\begin{figure}[ht]
\hspace*{0.5cm}\includegraphics[width=14cm]{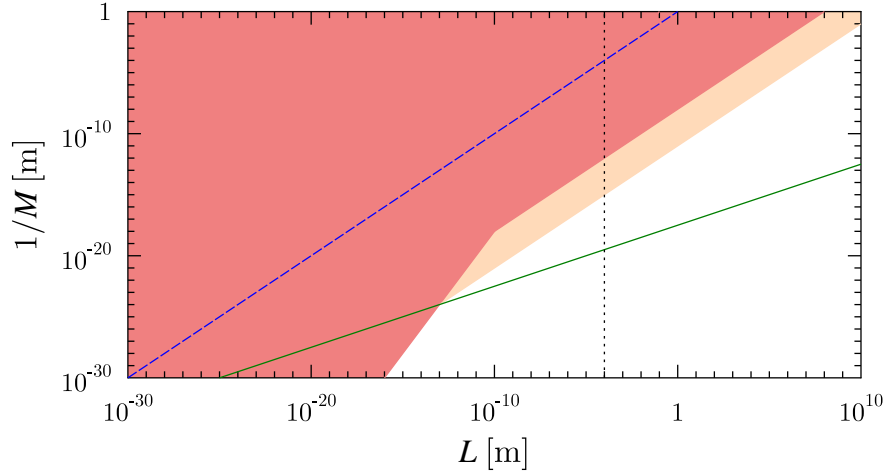} 

\caption{Summary of constraints. The coral (dark) shaded region is excluded. The peachpuff (light) shaded region indicates
the stronger constraint from photon decay with the ad-hoc assumption that the typical distance between 
defects increases with the cosmological scale factor. The dotted (black) line indicates the
length scale associated with the cosmological constant. The dashed (blue) line is the case $L M = 1$ and
the solid (green) line is the case $L M = 1/\epsilon$. \label{fig3}}
\end{figure}

The constraints on the parameter space from the previous sections are summarized in Figure \ref{fig3}

%..........................................................................
%..........................................................................

In hindsight, it is not entirely surprising that the bound from {\sc CMB} heating
is the strongest. The specific property, in fact the defining property, of 
the defects is that they violate
momentum conservation and, due to Lorentz-invariance, we cannot put a
hard cutoff on the magnitude of this momentum non-conservation. It is
only bounded by the properties of the incident particle and the larger
the spatial spread of the incident particle's probability distribution, the more
focused the defect and the larger the spread of the outgoing particle's
momentum. 

As we also found in \cite{own}, the present bounds are about 10 orders of magnitude
away from fully exploring the parameter range where the density of defects is
comparable to the cosmological constant, which in the cosmological setting
represents a natural range of parameters\footnote{It does not make sense to use values
of $1/M$ smaller than shown in Figure \ref{fig3}, because then we come into
the Planckian regime and the defects would be so dense they become the norm
rather than the exception.}. However, the constraints that we have considered here
are only estimates to gauge the promise of exploring space-time defects as
a signature for quantum gravitational effects. With a more sophisticated model
that takes into account background curvature, more of the existing cosmological
data could be analyzed. This would open the possibility of finding evidence for
space-time defects or at least deriving better constraints on their density.

\section{Discussion}

Let us first summarize the assumptions we have made that can in principle be
relaxed. We have
assumed that the defects don't carry quantum numbers, no spin or
gauge charges. We have restricted the study of phenomenological
consequences to the case where $M$ is larger than the electron mass.
We looked at the parameter range $\langle a^2 \rangle \sim 1$. The
latter assumption in particular could be modified. One could use $\Delta a^2$
to normalize the extension of the wave-packet into the third spatial
direction in a similar way as the perpendicular directions. We also remind ourselves that we have worked in the plane-wave
approximation, where $p_+ \gg \Delta p_+$. If this approximation
is not good, then the width of the defects can have a different dependence
on the momentum of the incident particle and the scaling of effects
might change.

As previously mentioned in \cite{own}, a certain case of nonlocal
defects effectively makes itself noticeable as a local defect. That will
be the case when the nonlocal translation can occur in both
directions between the same locations. In this case, a particle
that makes a nonlocal jump to another location will be replaced
at its point of departure with a different particle, making it
appear like a non-elastic scattering on a local defect. The problem
with this kind of scenario is that the probability for a particle to
appear at a certain location would depend on the total volume of 
spacetime, past and future, where it could have originated from. 
In this case, it is then impossible to say anything
about the interaction rates without first developing a
model for the generation of defects in a time-dependent background.

Finally, let us investigate the difference between the approach
discussed here and the one in \cite{Schreck:2012pf}. In the
model \cite{Schreck:2012pf}, interaction with the local defects is
mediated exclusively by a scalar field. The probability distribution of the
momentum that is assigned to the defect is not constrained by
a requirement similar to our requirements ${\bf p}\cdot {\bf k} = M^2$ and
${\bf k}^2 = a^2 M^2$.  As a consequence, Lorentz-invariance necessitates that the
defect be able to inject momenta from the full Lorentz-group, 
which is no longer normalizable. Thus
there arises the need to introduce a cutoff on the momentum
integration. While the model in \cite{Schreck:2012pf} offers an
concrete realization of coupling quantum fields to space-time
defects,  the need to eventually introduce a Lorentz-invariance violating
cutoff defeats the point of requiring a Lorentz-invariant
distribution and coupling to begin with. The more relevant difference
between the two models is however that we have here assumed
the coupling to appear as a contribution to the covariant
derivative and not as an independent interaction vertex.

These approaches are presently the only existing models to describe
space-time defects and the study of the effects is in its infancy. It is
possible, in fact likely, that elements of both approaches will turn
out to be necessary for the development of more sophisticated
models.

\section{Summary}

We have proposed a model for the scattering of particles on space-time
defects that induce a violation of energy-momentum conservation. In the plane wave-limit,
the energy-momentum non-conservation can become arbitrarily large due to
Lorentz-invariance, but
it remains bounded if one takes into account the finite widths of the
incident particle's wave-function. We have looked at various 
phenomenological consequences and estimated that the best
constraints come from energy deposited by decaying photons into 
the cosmic microwave background. 

\section*{Acknowledgements}
I thank Julian Heeck and Stefan Scherer for helpful discussions.


\begin{thebibliography}{99}
\small{ 
\bibitem{AmelinoCamelia:2008qg} 
  G.~Amelino-Camelia,
  {\it ``Quantum Gravity Phenomenology,''}
  Living Rev.\ Rel.\  {\bf 16}, 5 (2013)
  [arXiv:0806.0339 [gr-qc]].
  %%CITATION = ARXIV:0806.0339;%%
  %85 citations counted in INSPIRE as of 22 Aug 2013

%\cite{Hossenfelder:2009nu}
\bibitem{Hossenfelder:2009nu}
  S.~Hossenfelder and L.~Smolin,
  {\it ``Phenomenological Quantum Gravity,''}
Physics in Canada, Vol. 66 No. 2, Apr-June, p 99-102 (2010),
  arXiv:0911.2761 [physics.pop-ph].
  %%CITATION = ARXIV:0911.2761;%%

%\cite{Hossenfelder:2010zj}
\bibitem{Hossenfelder:2010zj} 
  S.~Hossenfelder,
  {\it ``Experimental Search for Quantum Gravity,''}
In {\it ``Classical and Quantum Gravity: Theory, Analysis and Applications,''} Chapter 5, Edited by V. R. Frignanni, Nova Publishers (2011),
  arXiv:1010.3420 [gr-qc].
  %%CITATION = ARXIV:1010.3420;%%
  %8 citations counted in INSPIRE as of 08 May 2013

\bibitem{own}
S.~Hossenfelder, {\it ``Phenomenology of Space-time Imperfection I: Nonlocal Defects''}

%\cite{Mattingly:2005re}
\bibitem{Mattingly:2005re} 
  D.~Mattingly,
  {\it ``Modern tests of Lorentz invariance,''}
  Living Rev.\ Rel.\  {\bf 8}, 5 (2005)
  [gr-qc/0502097].
  %%CITATION = GR-QC/0502097;%%
  %310 citations counted in INSPIRE as of 13 May 2013

%\cite{Kostelecky:2008ts}
\bibitem{Kostelecky:2008ts} 
  V.~A.~Kostelecky and N.~Russell,
  {\it ``Data Tables for Lorentz and CPT Violation,''}
  Rev.\ Mod.\ Phys.\  {\bf 83}, 11 (2011)
  [arXiv:0801.0287 [hep-ph]].
  %%CITATION = ARXIV:0801.0287;%%
  %267 citations counted in INSPIRE as of 13 May 2013


%\cite{Schreck:2012pf}
\bibitem{Schreck:2012pf} 
  M.~Schreck, F.~Sorba and S.~Thambyahpillai,
  {\it ``A simple model of pointlike spacetime defects and implications for photon propagation,''}
  arXiv:1211.0084 [hep-th].
  %%CITATION = ARXIV:1211.0084;%%

%\cite{Dowker:2003hb}
\bibitem{Dowker:2003hb} 
  F.~Dowker, J.~Henson and R.~D.~Sorkin,
  {\it ``Quantum gravity phenomenology, Lorentz invariance and discreteness,''}
  Mod.\ Phys.\ Lett.\ A {\bf 19}, 1829 (2004)
  [gr-qc/0311055].
  %%CITATION = GR-QC/0311055;%%
  %71 citations counted in INSPIRE as of 05 Jun 2013



%\cite{Bombelli:2006nm}
\bibitem{Bombelli:2006nm}
  L.~Bombelli, J.~Henson and R.~D.~Sorkin,
  {\it ``Discreteness without symmetry breaking: A theorem,''}
  Mod.\ Phys.\ Lett.\  A {\bf 24}, 2579 (2009)
  [arXiv:gr-qc/0605006].
  %%CITATION = MPLAE,A24,2579;%%


%\bibitem{Markopoulou:2007ha} 
 % F.~Markopoulou and L.~Smolin,
 % {\it ``Disordered locality in loop quantum gravity states,''}
 % Class.\ Quant.\ Grav.\  {\bf 24}, 3813 (2007)
 % [gr-qc/0702044].
  %%CITATION = GR-QC/0702044;%%

\bibitem{Weinberg}
S.~Weinberg, {\it The Quantum Theory of Fields, Volume I}, Cambridge University Press, Cambridge (1995).

%\cite{Goldhaber:2008xy}
\bibitem{Goldhaber:2008xy} 
  A.~S.~Goldhaber and M.~M.~Nieto,
  {\it ``Photon and Graviton Mass Limits,''}
  Rev.\ Mod.\ Phys.\  {\bf 82}, 939 (2010)
  [arXiv:0809.1003 [hep-ph]].
  %%CITATION = ARXIV:0809.1003;%%
  %74 citations counted in INSPIRE as of 09 Aug 2013

%\cite{Heeck:2013cfa}
\bibitem{Heeck:2013cfa} 
  J.~Heeck,
  {\it ``How stable is the photon?,''}
  Phys.\  Rev.\  Lett.\  111, {\bf 021801} (2013)
  [arXiv:1304.2821 [hep-ph]].
  %%CITATION = ARXIV:1304.2821;%%

%\cite{PrescodWeinstein:2009wq}
\bibitem{PrescodWeinstein:2009wq} 
  C.~Prescod-Weinstein and L.~Smolin,
  {\it ``Disordered Locality as an Explanation for the Dark Energy,''}
  Phys.\ Rev.\ D {\bf 80}, 063505 (2009)
  [arXiv:0903.5303 [hep-th]].
  %%CITATION = ARXIV:0903.5303;%%
}

%\cite{Jimenez:2013mga}
\bibitem{Jimenez:2013mga} 
  R.~Jimenez,
  {\it ``Beyond the Standard Model of Physics with Astronomical Observations,''}
  arXiv:1307.2452 [astro-ph.CO].
  %%CITATION = ARXIV:1307.2452;%%

%\bibitem{mirror} S.~Kuhr {\it et al}, {\it ``Ultrahigh finesse Fabry-Perot superconducting resonator,''}
%Appl.\ Phys.\ Lett.\ {\bf 90} (2007) 164101, [arXiv:quant-ph/0612138].





\bibitem{Beckmann} V.~Beckmann, Lecture Notes
of the Astrophysical Spring School, Carg\`ese/Corsica April 2006,
Retrieved July 12, 2013 at {\url{eud.gsfc.nasa.gov/Volker.Beckmann/school/download/Longair\_Radiation3.pdf}},
 


\end{thebibliography}
\end{document}